\documentclass[12pt]{article}

\usepackage[utf8]{inputenc}
\usepackage{amsmath, amssymb, amsfonts}
\usepackage{graphicx}
\usepackage{booktabs}
\usepackage{geometry}
\usepackage{setspace}
\usepackage{hyperref}
\usepackage{caption}
\usepackage{float}
\usepackage{newunicodechar}
\newunicodechar{≤}{\leq}

\hypersetup{
    colorlinks = true,
    linkcolor  = blue,
    citecolor  = blue,
    urlcolor   = blue,
    pdfborder  = 0 0 0
}

\title{Understanding statistics for biomedical research\\
through the lens of replication}

\author{
Huw Llewelyn\\[4pt]
Department of Mathematics, Aberystwyth University,\\
Penglais, SY23 3BZ, Ceredigion, United Kingdom\\[4pt]
\texttt{hul2@aber.ac.uk}
}

\date{}

\begin{document}
\maketitle

\newpage

\begin{abstract}
Clinicians and scientists are familiar with the concept of replication. However, P values that assess statistical significance, and confidence intervals, are not currently linked explicitly to the probability of replication. This makes P values and confidence intervals difficult to interpret in terms of replication. This paper describes a predictive model that allows P values to be interpreted in this way, provided that studies are assumed to be conducted impeccably with no errors, bias etc. For example, based on this model, if the sample size of a replicating study is very large, then the probability of replication with an effect size greater than zero in the same direction (i.e. same sign) is 1-P. However, if the replicating study sample size is the same as in the original, then the probability of getting the same statistically significant P value again is modest, as widely observed in empirical studies.
\\
This new approach models replication outcomes by using the sum of the variances of the original and replication studies obtained by convolving their two distributions. It does not incorporate for a second time the variance of the replicating study as in other models. This model is based on discretised rather than continuous values, which allow proper uniform prior probabilities. Under this predictive model, the probability of replication is conditional on the effect size, standard deviation, the sample size of the original study and the sample size of planned replication study. This model suggests that current modest replication rates are what should be expected in impeccably conducted studies. 
\\
 These replication rates can be calculated more simply from the P value alone of the original study and the expected P value in the repeat study. The replication probabilities from this simplified expression were well-calibrated against empirically observed replication frequencies in the Open Science Collaboration and also in Cochrane Database studies. In contrast, models that do incorporate for a second time the variance of the replicating study are not well calibrated. Calculations not based on convolved distributions and added variances will also underestimate required sample sizes. By helping non-statisticians to understand the link between P values and replication probabilities, this approach may provide a more accurate expectation of reproducibility in biomedical and behavioural research and provide more accurate sample sizes required to achieve specified P values.

\end{abstract}


\newpage

\section{Introduction}

 It was stated recently in a popular book written for a general readership that learning from data is “a bit of a mess” [1] and that confidence intervals have mystified generations of students [2]. The definition of a P value as the probability of an observed value, or some other more extreme hypothetical values, conditional on another single hypothetical value worries many that statistics may be impossible to understand fully. However, all students and scientists understand the idea of someone else replicating or confirming their own observations in clinics and labs, and the probability or expected frequency of this happening.

A high-profile statement by the ASA has tried to clarify the interpretation of P values.~[3, 4]. Nevertheless, it is still unclear how a P value is related to the probability of scientific replication, as reflected by the ongoing replication crisis affecting medicine and the social sciences~[5]. All this has been accompanied by persistent differences of opinion between Bayesians and Frequentists about how scientific data should be interpreted. In the Open Science Collaboration study, the average two-sided P value in 97 studies was 0.028, but only 36.1\% (95\% CI 26.6\% to 46.2\%) showed a two-sided P value of 0.05 or lower when each of the 97 studies was repeated~[6].

The estimation of probabilities and the replication of findings on which they are based are important for clinical observations during day-to-day medical practice as well as for studies that test scientific hypotheses. The uncertainty about replicating clinical observations reported by another clinician is a familiar daily experience for diagnosticians. An understandable model of replication is also important for physicians who are trying to assess how well the result of a study elsewhere would be replicated in their own clinical setting. The convention that measurements are based on interval scales in medicine and science may also suggest a way forward to understanding P values and confidence intervals based on Gaussian and other continuous model distributions. Discretising such distributions might also allow physicians and scientists to understand better the nuances of Bayesian and Frequentist statistics.

\section{A directly estimated probability of an outcome without Bayes rule}\label{sec2}

If 100 patients had been in a double-blind cross-over randomized controlled trial and 58 out of those 100 individuals had a BP that was higher on control than on treatment, then knowing only that an individual was one of those in the study, then, knowing only that an individual was one of those in the study, a diagnostician’s probability—conditional on the entry criterion—of that individual having a BP difference greater than zero would be about 58/100 = 0.58. This would be a probability estimated directly from observed outcome frequencies, without first estimating prior probabilities, likelihood distributions, and then applying Bayes’ rule.

\subsection{Probabilities conditional on numerical results}\label{subsec1}

If in a study on many subjects the average BP difference between a pair of observations on treatment and control was 2\,mmHg, and the standard deviation (SD) of the differences was 10\,mmHg, then in Figure~1 the area under the broad bell-shaped Gaussian distribution above 0\,mmHg (i.e.\ 0.2 = (2–0)/10 standard deviations below the mean of 2\,mmHg) would contain CDF(0.2) = 58\% of the total area. (CDF means the cumulative distribution function.) This again would suggest a probability of a BP difference greater than zero that is estimated directly without first estimating prior probabilities, likelihood distributions, and then applying Bayes’ rule. From this we again see that a randomly selected study individual will have a probability of 0.58 of having a BP difference greater than zero.

\begin{figure}[htbp]
    \centering
    \includegraphics[width=1\linewidth]{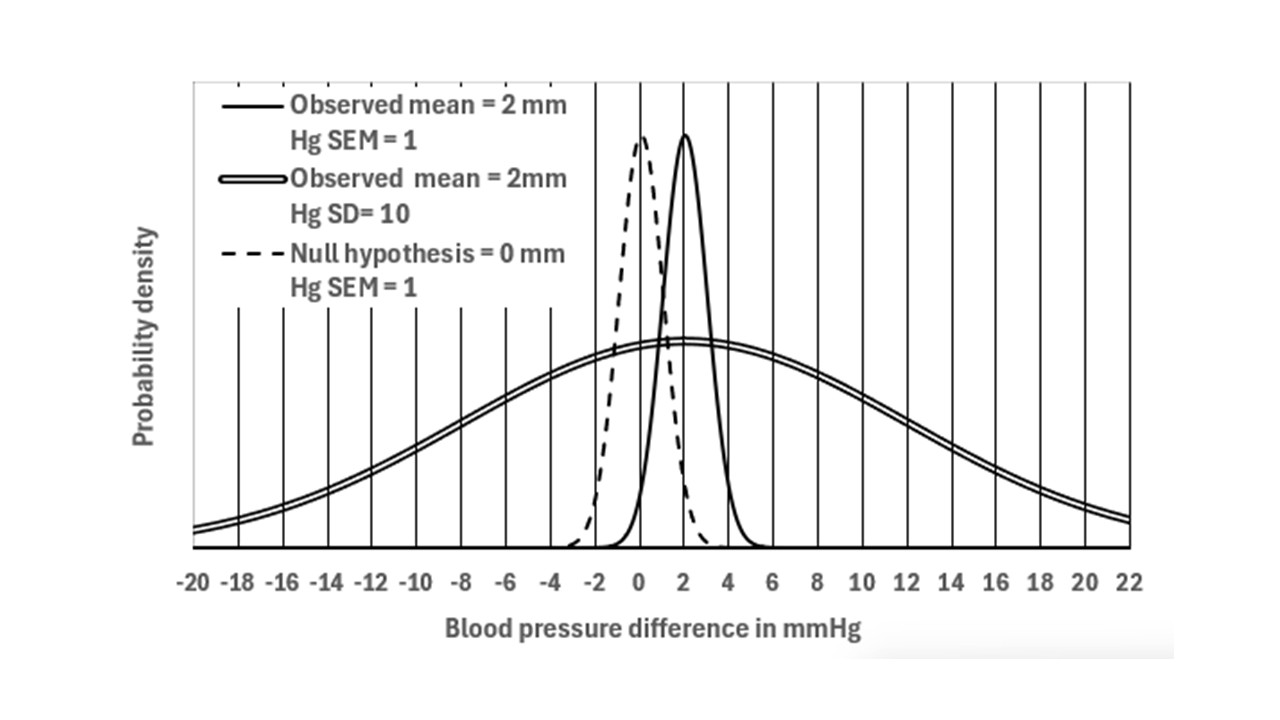}
    \caption{The distributions of blood pressure differences in a cross-over RCT}
    \label{fig:placeholder}
\end{figure}

\section{Distribution of possible means of different studies}\label{sec2}

As the standard deviation (SD) of the distribution was 10\,mmHg and the number of observations was 100, the standard error is found by dividing the SD by the square root of the number of observations: $10/\sqrt{100} = 1$. This means that, having obtained a mean difference of 2\,mmHg in the first study, if we repeated the study of 100 observations a very large number of times, then the probability of one of these many subsequent studies showing a mean difference greater than zero (which is 2 SEMs away from 2\,mmHg) is CDF(2) = 0.97725. It follows that the probability of one of the many study means being 0 or less is $1 - 0.97725 = 0.02275$. It also follows that if the mean difference of the first study had been 0\,mmHg then the probability of a repeat study of 100 observations showing a 2\,mmHg difference or something more extreme would also be 0.02275.  

\subsection{Discretising the Gaussian distribution}

The Gaussian distribution is smooth because it is a continuous function. This is mathematically convenient in some ways but causes difficulty when trying to interpret results using Bayes’ rule because it creates mathematical ambiguities. This can be overcome by converting the Gaussian distribution from a smooth function into a histogram with very narrow discrete columns (e.g.\ of 0.01\,mmHg width). This process is called “discretisation”. It corresponds to how clinicians and scientists record measurements not as single values but as numerical intervals, naming the interval according to the higher or lower numerical value making up the interval. For example, a BP reported as 120\,mmHg on a millimetre scale implies that it was greater than or equal to 120\,mmHg but less than 121\,mmHg.

It was estimated that when the first observed mean was 2\,mmHg the probability of a true mean being zero or less was 0.02275. It follows that the probability of the true mean being minus 0.01\,mmHg or less would be $1 - \text{CDF}(2 - (-0.01)) = 1 - \text{CDF}(2.01) = 0.02222$. This means that the probability of these study results being between $-0.01$\ and 0 mmHg, would be $0.02275 - 0.02222 = 0.00053$. Similarly, if the first study result had been 0\,mmHg, the probability of a repeat study result being 2.01\,mmHg would be $1 - \text{CDF}(2.01 - 0) = 0.02222$. This means that the probability of the next study being between 2\,mmHg and 2.01\,mmHg would be $0.02275 - 0.02222 = 0.00053$. If we repeated these calculations from $-4$\,mmHg to $+8$\,mmHg for a mean of zero and 2\,mmHg, we would obtain the two narrow distributions in the centre of Figure~1, each with $1200 - 1 = 1199$ columns of width 0.01\,mmHg. This requires 1199 rows on a spreadsheet.

\subsection{Likelihood distributions}

A Gaussian distribution can also be used to model the possible results of making 100 random selections from a population of studies with a true mean difference of 0\,mmHg (used as a null hypothesis for differences) when the SEM of the selected results is 1\,mmHg. This is a continuous likelihood distribution with values that represent likelihood densities, which, when added, do not sum to one. In this situation, the probability of getting a result of 2\,mmHg or more would also be 0.02275, as above. (This would, of course, correspond to a P value of 0.02275.) This value would normally be calculated from a Z score, or standard normal deviate, of 2. For a result of 2.01\,mmHg or more, it would be 0.02222, as shown already. Finally, the probability of getting a result between 2 and 2.01\,mmHg would also be 0.00053 given a null hypothesis of zero.

If we label the interval of 0 to $-0.01$\,mmHg as being 0 and we label the interval of 2 to 2.01\,mmHg as 2\,mmHg, then the probability of a hypothetical true mean of 0\,mmHg conditional on observing a mean of 2\,mmHg is 0.00053, and the probability of observing a mean of 2\,mmHg conditional on a hypothetical true mean of 0\,mmHg is also 0.00053. According to Bayes’ rule, if we know already that the two conditional probabilities are equal it follow that the prior probability of the true mean of zero and an observed mean of 2\,mmHg must be equal. If we specify that the finite range of values that we are going to deal with is from $-4$\,mmHg to $+8$\,mmHg, this results in $12 \times 100 - 1 = 1199$ equal intervals or “bins”. Each would have the same, or “uniform”, prior probability of about $1/1199 = 0.00083$, summing to 1. In a spreadsheet, they would be displayed on 1199 rows and graphically as a flat curve 0.00083 above the baseline.

Currently, it is assumed that as uniform prior probabilities have an infinite range, this renders them “improper” because they cannot sum to one, and that the prior probability of any particular continuous value is zero. Instead, it should be possible to assume in general that the Gaussian probability curve is made up of extremely small intervals, each labelled $\delta$ (as in differential calculus), with a finite range from $-1/\delta$ to $+1/\delta$ (with $-4$ to $+8$\,mmHg in the above example being particular values of the former). This would mean that under these assumptions flat, uniform prior probabilities of the parameter and statistic would be “proper”, so that we can assume that
\[
p(\text{Parameter} \mid \text{Statistic}) = p(\text{Statistic} \mid \text{Parameter}),
\]
and calculations could be made using closed formulae. This latter approach will be taken with most of the following calculations. Discretised bins of 0.01\,mmHg in Excel were used to create Figures~1 and~2.

\subsection{Different Gaussian distributions of means}\label{sec2}

The distribution of possible means can represent three different situations:\\

(A) The distribution of probabilities of all possible true means (“true” after an infinite number of observations) conditional on the observed mean and a standard error of the mean (SEM) of a limited number of measurement results (e.g., as represented by a distribution with a mean at 2\,mmHg in Figure~1). The prior distribution of all possible hypothetical means is conditional on the sample space (i.e., before the nature of any future study is known, so that a Bayesian prior distribution cannot yet be estimated). Based on this uniform distribution conditional on the sample space and Bayes’ rule, the converse likelihood distribution of the observed mean conditional on all possible hypothetical means can be assumed to have the same form as the corresponding probability distribution (and can be made identical by normalisation).

(B) The likelihood distribution of possible observed means conditional on a single hypothetical true mean, assuming  prior probability of all hypothetical true means as in (A), so that the SEM and underlying mathematical model are the same as in (A) (e.g., the likelihood distribution with a mean at a null hypothesis of 0\,mmHg as in Figure~1).

(C) The likelihood distribution of a unique observed mean conditional on each of the possible hypothetical means, again assuming that the SEM is the same as in (A) (e.g., as represented by a distribution with a mean at 2\,mmHg in Figure~1).

\subsection{The null hypothesis}\label{subsec1}

Based on situations (A) and (B), the tail area of the distribution in Figure~1 more extreme than the observed mean (e.g., $\geq 2$\,mmHg in Figure~1) will be the same as the tail area of the distribution more extreme than the null hypothesis (e.g., $\leq 0$\,mmHg in Figure~1). This is because both distribution tails are the same number of SEMs away from their respective distribution means. Therefore, the P value will also be equal to the probability of the true mean being more extreme than the null hypothesis (e.g., less than 0\,mmHg) conditional on the observed mean (e.g., 2\,mmHg). Thus, the probability of the true mean being less extreme than the null hypothesis (e.g., $>0$\,mmHg) conditional on the observed mean will be equal to $1$ minus the latter, and $1$ minus the P value. Accordingly, the likelihood of getting the observed mean difference of 2\,mmHg or something more extreme (i.e., $>2$\,mmHg) conditional on the null hypothesis in Figure~1 is the P value of 0.02275. The probability of the true difference being zero or lower is also 0.02275, and the probability of it being above zero is 0.97725.

\subsection{Why Bayesian prior probabilities are not conditional on the sample space}\label{subsec1}

The Bayesian prior probability is different from the uniform prior probability conditional on the sample space described above. The Bayesian prior is estimated not before but after designing the study, by conducting a thought experiment based on personal experience and reading the literature to estimate what the distribution of possible results might be in an actual study, conditional on background knowledge. Any subsequent observed result will therefore be independent of the prior estimated result, so that the former is not a subset of the latter, which is an essential condition for Bayes’ rule and the presence of a sample space. By contrast, the possible “true” and observed means will both be subsets of the sample space.

Therefore, the Bayesian prior distribution can be regarded as a posterior distribution formed by combining  prior distribution conditional on the sample space with an estimated likelihood distribution of the thought-experiment study result (or pilot study result) conditional on all possible true values. Each of these likelihoods is then multiplied by the likelihood of observing the actual study result conditional on all possible true results. These products are then normalised to give the Bayesian posterior probability of each possible true result, conditional on the combined evidence of the result of the Bayesian thought experiment and the actual study result.

\subsection{The baffled student}\label{subsec1}

We could now explain to a student (based on the above considerations and assumptions) that a one-sided P value of 0.02275 is numerically the same as the probability of the true study mean falling into a range beyond the null hypothesis, conditional on the observed data, methods, and mathematical models used. This also implies that there is a probability of 0.97725 that the true mean does fall into a range on the same side as the observed mean. In some cases, the range of interest might be some other interval between two thresholds (e.g., a BP difference between 0\,mmHg and 3.96\,mmHg or between 1\,mmHg and 3\,mmHg). This often happens, of course, when an investigator wishes to test the hypothesis that there is little difference between the outcomes of treatment and control.

In this case, there would be a probability of $1 - 0.0228 - 0.0228 = 0.9544$ that the true result mean would fall within the range of interest of 0\,mmHg to 3.96\,mmHg, or a probability of 0.683 that the true result mean will fall within the range of interest of 1\,mmHg to 3\,mmHg, or a probability of 0.5 that it will fall within the range of interest above 2\,mmHg. It follows that there is a probability of 0.95 that the true result lies within the 95\% confidence interval.

\section{Replication}

\subsection{A single result}
It could be assumed form the foregoing assumptions that when a clinician is given a test result with a standard deviation, this can be used to predict with a probability of 0.95 that the true result will fall within two standard deviations of the result. This could be based on the assumption that the scale on which the observation is made and the scale of possible true values are the same. Because of this the prior probability of any true discretized value and the prior probability of the observed value are the same and uniform (see section  3.1, Discretising the Gaussian distribution). Therefore, if 97.5\% of the above range centered around the observed single result was above a diagnostic threshold, then we might infer that there was a probability of 0.975 that the true result was above that threshold. 

If the first test result were not used to estimate the probability of a true result being above a threshold, but the probability of a second test result also being above the threshold, this would represent a probability of replicating the result of the first test. This probability would be lower than the probability of a true result being above the threshold. This is because there are two errors involved. The first is in estimating the probability distribution of the possible true results, the second error being when estimating the probability distribution of the second result conditional on each of the possible true results. The distribution of the second result is obtained by convolving the distribution of the true result with the distribution of the second result conditional on each possible true result.   

The first error is represented by the bell-shaped distribution of the possible true result conditional on the first test result. For each possible true result on the first bell shaped distribution, there is an equal error in estimating the result of the next test result, also based on a bell-shaped distribution. For the same test being repeated, these two distributions will have the same variance. Therefore, the overall variance based on their convolution is twice the variance for a single test result.

To summarise, the process of predicting the result of a second test conditional on the first can be broken down into two stages. The first stage is to use the result of the original completed study, under the assumed uniform prior distribution, to estimate the probability distribution of the possible “true” mean—that is, our probability distribution over the mean that would be approached if the original study were continued to a very large number of observations. The second stage is to estimate the probability distribution of the possible means of a replicating study conditional on each possible value of this “true” mean. Integrating over these conditional distributions gives the predictive distribution of the replication-study mean. This represents a convolution of the two distributions, modelled by adding the variances of two distributions. However, the facts on which the probability of replication is based are the mean and variance of the test result.

I should be emphasised that the convolution of the two distributions represents the distribution of the possible results of the second replicating observations conditional on the mean and standard deviation of the original set of observations. It may be tempting to postulate the presence of a single possible replication study result with a mean equal to the original mean and to use this to establish a replication threshold. However, this distribution is already been taken into consideration as one of the many distributions represented by the convolved distribution and will result in exaggerating the probability of replication.

\subsection{The mean of many study results}
An analogous process can be applied to scientific studies. In this situation, instead of predicting the range within which a single value of the repeat test falls, we now wish to predict the mean of a set of results obtained from a second replicating study yet to be done conditional on the mean and sample variance of the first study that has been done. The process of predicting the mean of a second replicating study conditional on the original study result can similarly be broken down into two stages. The first stage is to use the result of the original completed study to estimate the probability distribution of the possible “true” means (i.e.\ the possible means that would be discovered if the first study had been continued until there was a vast number of observations). 

The second stage is to estimate the probability distribution of the possible means of a replicating study conditional on each possible value of this “true” mean. Integrating over these conditional distributions gives the predictive distribution of the replication-study mean. This represents a process of convolution that creates a new distribution of the future means of the replication study, conditional on the result (means and SEM) of the original study. If the second replicating study has a different planned sample size, then its variance will be different. Therefore the variance of the convolved distribution will be  equal to the sum of the variances of the first and second study distributions. 

We cannot know or even estimate the P value of the second study per se, as we do not know its mean, but only the SEM and sample variance of the distribution of its possible means from the convolved distribution. We can use this to find the range of results in the new convolved distribution for which the P value will be 0.025 or less with respect to the original observed mean and SEM, and estimate the probability distribution of this happening. This is known as the probability of statistically significant replication. We can also estimate the greater probability that the result of the replicating study will be greater than zero, irrespective of statistical significance, in the same direction (AKA 'the same sign replication').

\subsection{The combined variance of two independent studies}\label{subsec1}

To simplify matters, we now use a slightly different example where the observed mean difference is 1.96\,mmHg instead of 2\,mmHg, but the standard deviation is still 10\,mmHg and the SEM is still 1 based on 100 observations. However, if we postulate a distribution of differences for the result of the second study, also with an SEM of 1, added to the distribution of true mean values of the first study (the convolution of the two distributions with the same variance), then the combined convoluted distribution will have a variance of $1 + 1 = 2$, and the SEM will be $\sqrt{2} = 1.414$\,mmHg (see Figure~2).
\
\begin{figure}
        \centering
        \includegraphics[width=0.9\linewidth]{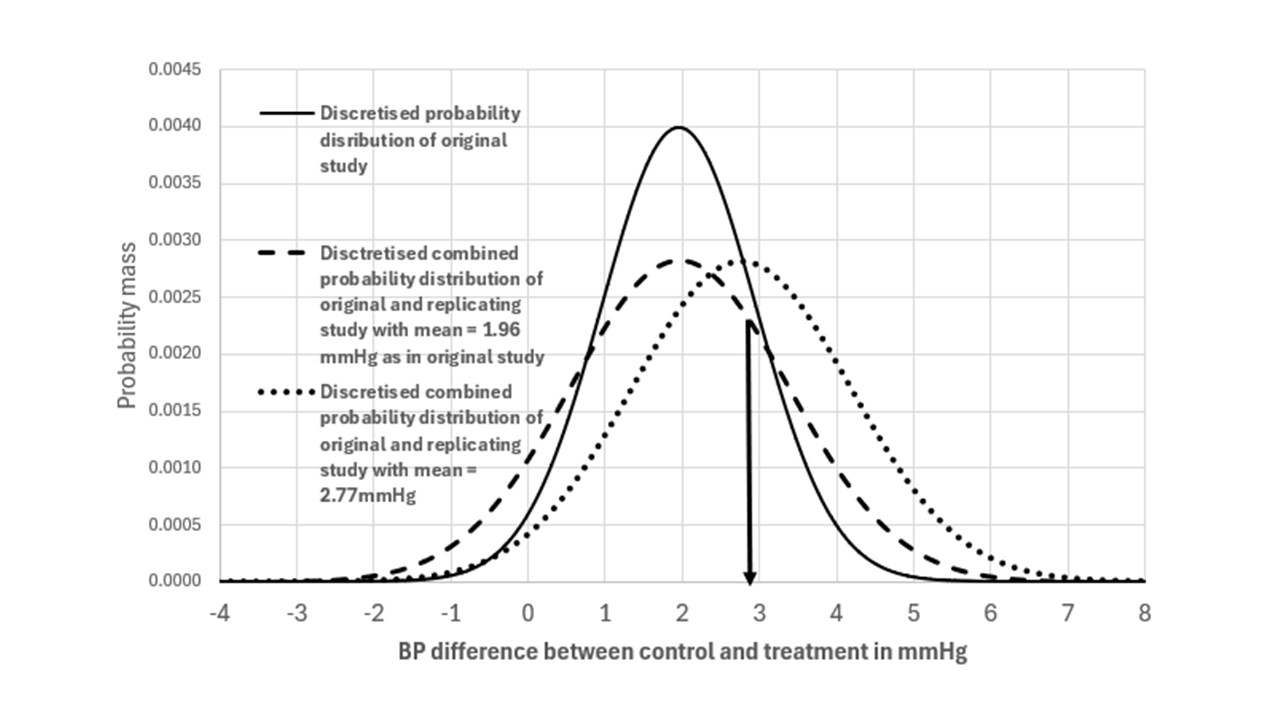}
        \caption{The distribution of blood pressure differences when the results of one study are followed by the results of an identical study.}
        \label{fig:placeholder}
    \end{figure}

To achieve a combined result that provides a P value of $\leq 0.025$ again, the many possible replicating study results must be at least 2.77\,mmHg (as indicated by the arrow in Figure~2). This is because the standard error of the convolved predictive distribution is 1.414 mmHg and the result must be 1.96 SEMs away from the null hypothesis of zero. Therefore, the repeat result must be at least $1.414 \times 1.96 = 2.77$\,mmHg away from zero, which is $2.77 - 1.96 = 0.81$\,mmHg away from the observed mean of 1.96\,mmHg. This corresponds to a Z score of $-(0.81/1.414) = -0.573$. The area of the distribution with mean 1.96\,mmHg in Figure~2 to the right of the arrow is therefore 28.3\% of its total area. 

Note that the discretised probability distribution of a replicating study with the same mean and SEM as the original study will be identical to that of the original study as displayed in Figure-2. If the threshold for replication is based on this single replicating distribution, the black arrow will be moved to the left and placed at 1.96 mmHg instead of 2.77 mmHg. This also corresponds to the mean and median of the convolved distribution so that 50\%  of the convolved distribution will lie to its right. This corresponds to a probability of significant replication of 0.5 used in some other approaches to estimating the probability of replication success, which appears to create a bias. This probability of 0.5 compares to a probability of 0.283 corresponding to the area of 28.3\% to the right of the existing arrow at 2.77 mmHg. It is the latter rationale of using the SEM of the convolved distribution to place the replicating threshold (e.g. at 2.77 mmHg) that is applied here.

The foregoing discussion describes a numerically intuitive understanding of random selection modelled by uniform prior probability distributions and the convolution of probability distributions that represents replication. This numerical rationale will now be formalised in mathematical notation.

\section{The argument in terms of mathematical notation}

Let $P_1$ be the one-sided P value from the first completed study (e.g.\ 0.025) and $P_2$ the one-sided P value of the planned replicating study. ($P_2$ is therefore unknown as it is in the future; we can only estimate its probability of being less than or equal to a threshold, e.g.\ $\leq P_3 = 0.025$.) $P_3$ is therefore the specified desired upper threshold of the one-sided P value in the second replicating study. $b$ is the observed raw effect size in the first study (e.g.\ 1.96\,mmHg), $s$ is the standard deviation (e.g.\ 10\,mmHg), $n_1$ is the sample size of the first completed study (e.g.\ 100), and $n_2$ is the planned sample size of the second replicating study (e.g.\ 100). In general terms (for proof, see Appendix~1):
\[
p(P_2 \leq P_3 \mid b, s, n_1, n_2)
 = \Phi\left(
 \frac{b}{\sqrt{ \left( \frac{s}{\sqrt{n_1}} \right)^2
 + \left( \frac{s}{\sqrt{n_2}} \right)^2 }}
 + \Phi^{-1}(P_3)
 \right)
\tag{Equation 1}
\]
and by applying the above example data:
\[
p(P_2 \leq 0.025 \mid b, s, n_1, n_2)
 = \Phi\left(
 \frac{1.96}{\sqrt{
 \left( \frac{10}{\sqrt{100}} \right)^2
 + \left( \frac{10}{\sqrt{100}} \right)^2 }}
 + \Phi^{-1}(0.025)
 \right)
 = 0.283
\tag{Equation 2}
\]

However, when $n_1 = n_2$, the above equation simplifies to:

\[
p(P_2 \leq 0.025 \mid b, s, n_1, n_2)
 = \Phi\left(
 \frac{1.96}{\sqrt{
 \left( \frac{10}{\sqrt{100}} \right)^2 \cdot 2 }}
 + \Phi^{-1}(0.025)
 \right)
 = 0.283
\tag{Equation 3}
\]

which in general terms is:

\[
p(P_2 \leq P_3 \mid b, s, n_1, n_2)
 = \Phi\left(
 \frac{b}{\sqrt{
 \left( \frac{s}{\sqrt{n_1}} \right)^2 \cdot 2 }}
 + \Phi^{-1}(P_3)
 \right)
\tag{Equation 4}
\]

which rearranges and simplifies to (see Appendix~2):

\[
p(P_2 \leq P_3 \mid P_1)
 = \Phi\left(\Phi^{-1}(P_3) -\Phi^{-1}(P_1)/\sqrt{2}\right)
\tag{Equation 5}
\]
\\
This shows that the probability of replication with $P \leq 0.025$ one-sided (or $P \leq 0.05$ two-sided) with a study of the same sample size can be calculated from the P value alone of the first study.

\begin{table}[h!]
\centering
\renewcommand{\arraystretch}{1.2}
\begin{tabular}{|c|c|c|c|}
\hline
P value $(P_1)$ & Cochrane Database & Goodman et al.\ & Equation 5 \\
\hline
0.5 & 0.11 & 0.18 & 0.069 \\
\hline
0.3 & 0.16 & 0.26 & 0.110 \\
\hline
0.1 & 0.23 & 0.41 & 0.213 \\
\hline
0.05 & 0.29 & 0.50 & 0.283 \\
\hline
0.03 & 0.34 & 0.56 & 0.335 \\
\hline
0.01 & 0.44 & 0.67 & 0.445 \\
\hline
0.005 & 0.50 & 0.73 & 0.510 \\
\hline
0.001 & 0.64 & 0.83 & 0.643 \\
\hline
\end{tabular}
\caption{Column 1 shows various P values from an analysis of a large number of publications in the Cochrane Database. Column 2 shows the corresponding frequencies of replication with $P \leq 0.05$ again. Column 3 shows the predicted probability of replication derived from Goodman's work, and Column 4 the probability of replication predicted by Equation~5.}
\end{table}
Table~1 shows P values and corresponding frequencies of replication with P values $\leq 0.05$ two-sided obtained indirectly from the Cochrane Database~[7]. The authors pointed out that, according to their calculations, these estimated frequencies of replication were lower than expected~[7, 8]. For example, when the P value was 0.05 (two-sided), the Cochrane frequency was 0.29. This value had been expected to be about 0.5 based on calculations derived from Goodman’s earlier work in 1992~[9]. However, the Cochrane Database frequencies were matched closely by those based on Equation~5. For example, when $P = 0.05$ two-sided, the calculated frequency of replication from Equation~5 is 0.283, while from the Cochrane Database it was 0.29. This suggests that Equation~5 was based on realistic assumptions (e.g.\ uniform prior probabilities, adding variances of the original and replicating studies, etc.), but at least one of the assumptions made when applying Goodman's work was not realistic.

The data from the Open Science Collaboration study~[6] were based on pairs of very similar original and replicating studies. The average P value from the original 97 studies was 0.028 (two-sided), and for this average P value, 35/97 = 36.1\% (95\% CI 26.6\% to 46.2\%) showed a two-sided P value of 0.05 or lower when repeated. According to Equation~5, this proportion was predicted as 34.2\%. This slight difference could be explained by the 95\% confidence interval of 26.6\% to 46.2\% for the observed proportion of 36.1\%. Nevertheless, the frequencies were very close to those predicted by Equation~5, as were those from the Cochrane Database. The “Many Labs” study also suggested that frequencies of replication were lower than expected from expressions derived from Goodman’s work.

\subsection{The effect of the replicating study's sample size being greater (including when approaching infinity) than that of the original study}

Equation~5, which is based on P values, assumes that the sample size in the replicating study is the same as in the original study. To examine the effect of unequal sample sizes, we have to use Equation~1. Equation~2 showed that when $b = 1.96$, $s = 10$, and $n_1 = n_2 = 100$, then the probability of replication with $P \leq 0.025$ is 0.283. If, for the sake of argument, the replicating study’s sample size is $n_2 = 200$, then the probability of replication with $P \leq 0.025$ would be 0.360:

\[
p(P_2 \leq 0.025 \mid b, s, n_1, n_2)
= \Phi\left(
 \frac{1.96}{\sqrt{
 \left( \frac{10}{\sqrt{100}} \right)^2
 + \left( \frac{10}{\sqrt{200}} \right)^2 }}
 + \Phi^{-1}(0.025)
 \right)
 = 0.360
\tag{Equation 6}
\]

Senn pointed out that there is no need for the replicating study to have the same sample size as the original~[11]. Theoretically, a replicating study with an extremely large (approaching infinite) sample size would uncover the truth, so any non-replication would be due solely to the stochastic characteristics of the original study. If we replace 200 with a very large number approaching infinity (i.e.\ $n_2 \rightarrow \infty$) in Equation~6, then
\[
\left( \frac{10}{\sqrt{n_2 \rightarrow \infty}} \right)^2 = 0,
\]
so we obtain a probability of replication of 0.5:

\[
p(P_2 \leq 0.025 \mid b, s, n_1, n_2)
= \Phi\left(
 \frac{1.96}{\sqrt{
 \left( \frac{10}{\sqrt{100}} \right)^2
 + \left( \frac{10}{\sqrt{n_2 \rightarrow \infty}} \right)^2 }}
 + \Phi^{-1}(0.025)
 \right)
\]
\[
= \Phi\left(
 \frac{1.96}{\sqrt{
 \left( \frac{10}{\sqrt{100}} \right)^2 }}
 + \Phi^{-1}(0.025)
 \right)
 = \Phi\left( -\Phi^{-1}(0.025) + \Phi^{-1}(0.025) \right)
 = 0.5
\tag{Equation 7}
\]

In other words, when we have a P value of 0.025 one-sided (or 0.05 two-sided), then the probability of replication with second study of near-infinite sample size (which would “discover the truth”) is 0.5, numerically the same as that estimated earlier by Goodman for a repeated study of equal size~[9]. However, if the P value were 0.0025, the probability of such replication would be 0.8. It seems more appropriate to estimate probabilities of replication by “taking a long view” of what we would expect if a large community of scientists repeated the study and pooled their results to obtain a “true” result based on a replicating study with a near-infinite sample size.

\subsection{P value or “same sign” replication}

It should be emphasised that the foregoing discussion concerns “P value replication,” in which the result of interest is the P value when the study is repeated. However, another form of replication is “same sign replication,” where instead of focusing on the P value, we focus on whether the difference between treatment and control is in the same direction in the replicating study. For example, if treatment was better than control in the completed study (e.g.\ a positive result greater than the null hypothesis of zero difference), will this also be the case in the replicating study?

Killeen proposed $P_{\mathrm{rep}}$ as an alternative to the P value. $P_{\mathrm{rep}}$ is the probability of obtaining an effect of the same sign as that found in the original experiment if the study were repeated in exactly the same way with the same sample size~[12]. This was based on a calculation similar to Equation~5 but replacing $P_2 \leq 0.025$ with $Z_{\mathrm{repl}} > 0$ and omitting $+\Phi^{-1}(0.025)$, giving:
\[
P_{\mathrm{rep}}
 = p(Z_{\mathrm{repl}} > 0 \mid P_1)
 = \Phi\left( -\Phi^{-1}(P_1) / \sqrt{2} \right)
\tag{Equation 8}
\]

When $P_1$ (the P value for the original study) is 0.025 (one-sided), then $P_{\mathrm{rep}}$ is 0.917 for seeing an effect size $>0$ in the same direction again when a study is repeated with the same sample size. However, if the second (replicating) study has a theoretical sample size of infinity, then when the P value is 0.025, the $\sqrt{2}$ in Equation~8 is replaced by $\sqrt{1}$ in Equation~9 because the variance of the second study is zero. The probability of replication is then 0.975:

\[
P_{\mathrm{rep}}
 = p(Z_{\mathrm{repl}} > 0 \mid P_1)
 = \Phi\left( -\Phi^{-1}(P_1) / \sqrt{1} \right)
\tag{Equation 9}
\]

We have therefore come full circle and can conclude that the P value suggests that if the study had been designed and performed impeccably and continued until we had an infinite number of samples, then the probability of a result of the same sign (e.g.\ greater than the null hypothesis) would be 0.975~[13]. We obtain the same result if the original study is not continued but instead repeated in a replicating study of infinite sample size. However, if we repeat the study with the same sample size, the probability of getting a same-sign result is 0.917. Furthermore, if it is repeated with the same sample size, the probability of getting $P \leq 0.025$ again is only 0.283.

It is only probabilities of replication with studies of similar sample sizes that can be verified empirically. This was done by the Open Science Collaboration [6]. van Zwet et al used a less direct approach based on assumptions about meta-analytic modeling [7, 8]. The latter verifications support the validity of Equations 1 to 9, suggesting that the predictions of what would happen with a replicating study of near-infinite sample size are also valid. This also supports the validity of using an uniform prior probability when modelling random selection. In almost all other cases (e.g. in medical diagnosis) it would not be appropriate to assume uniform prior probabilities (often called 'base rate neglect').

\subsection{Power calculations and estimating the probability that a study would have $P \leq 0.05$ two-sided}

Power calculations are based on the concept of detecting some fixed effect size and therefore give rise to “likelihood” probabilities. However, the foregoing Equations~1 to~9 estimate probabilities of obtaining a study result by first estimating the probability distribution of possible true effect sizes. An interesting aspect of Equation~7 is that it is the inverse of a power calculation. If we rearrange Equation~7, we obtain the sample size required for a power of 50\%:

\[
\left( 10 \times \frac{\Phi^{-1}(0.5) - \Phi^{-1}(0.025)}{1.96} \right)^2 = 100
\tag{Equation 10}
\]

In general terms (where $n$ is the required sample size, SD is the standard deviation, $1-\beta$ is the required power, and $b$ is the estimated raw effect size), the equation is:

\[
n = 
\left(
\mathrm{SD} \times 
\frac{\Phi^{-1}(1-\beta) - \Phi^{-1}(\alpha/2)}{b}
\right)^2
\tag{Equation 11}
\]

If a thought experiment had led us to imagine a theoretical pilot study with an SD of 10, a raw effect size of 1.96, and a sample size of 50, then the probability of replicating this result with $P \leq 0.025$—taking into account the distribution of possible true results convolved with the distribution of $b_{\mathrm{repl}}$ from Equation~1—would be:

\[
p(P_2 \leq 0.025 \mid b, s, n_1, n_2)
= \Phi\left(
\frac{1.96}{\sqrt{
\left( \frac{10}{\sqrt{50}} \right)^2
+ \left( \frac{10}{\sqrt{50}} \right)^2 }}
+ \Phi^{-1}(0.025)
\right)
= 0.164
\tag{Equation 12}
\]

In order to achieve a power of 80\% according to Equation~10, the required sample size is 204.3, rounding up to 205. When the thought-experiment study has a sample size of 205, the probability of replication in the planned study is still only 0.510. When the sample size of the thought experiment is increased to 409, the probability of obtaining $P \leq 0.025$ becomes 0.8.

There do not appear to be data available in the literature to support these estimated frequencies of obtaining $P \leq 0.025$ again when a planned study has been completed. The only data available relate to replication studies in which the completed planned study is repeated. If we regard the variance of the distribution of $b_{\mathrm{repl}}$ in the first real study as $v_1 + v_1 = 2v_1$ (see Table~2), and we repeat that study with the same sample size to create a replication study, then the combined variance will be $v_1 + v_1 + v_1 = 3v_1$. We can therefore modify Equation~2 to represent three variances, but with a sample size of 205 based on a power of 80\%:

\[
p(P_2 \leq 0.025 \mid b, s, n_1, n_2)
= \Phi\left(
\frac{1.96}{\sqrt{
\left( \frac{10}{\sqrt{205}} \right)^2 \cdot 3 }}
+ \Phi^{-1}(0.025)
\right)
= 0.367
\tag{Equation 13}
\]

\begin{table}[h!]
\centering
\renewcommand{\arraystretch}{1.2}
\begin{tabular}{|c|c|c|c|c|}
\hline 
Sample size required for cross-over RCT & 100 & 205 & 409 & 613 \\
\hline
Statistical Power (based on one variance = $v$) & 0.500 & 0.801 & 0.977 & 0.998 \\
\hline
Prob.\ of $P \leq 0.025$ in 1st study (2 variances = $2v$) & 0.283 & 0.510 & 0.800 & 0.929 \\
\hline
Prob.\ of $P \leq 0.025$ in 2nd study (3 variances = $3v$) & 0.207 & 0.367 & 0.629 & 0.800 \\
\hline
Total sample size required for parallel RCT & 400 & 820 & 1636 & 2452 \\
\hline
\end{tabular}
\caption{For various sample sizes in the planning stage when the expected mean is 1.96\,mmHg and the standard deviation is 10\,mmHg, the power and probability of replication with $P \leq 0.025$ one-sided (or $P \leq 0.05$ two-sided) again when the replicating study is done with the same sample size (based on Equation~13).}
\end{table}

In order to obtain a probability of 0.8 (rather than 0.367) of obtaining $P \leq 0.025$ in the replicating study, we would need to replace the predicted sample size of 205 in Equation~13 with 613 (see Table~2). This would give a probability of 0.929 of obtaining $P \leq 0.025$ in the first study based on a “power” of 0.998. In order to obtain a probability of $P \leq 0.025$ in the first study, we need a sample size twice that of 205—namely 409. A sample size of 100 provides only a probability of 0.283 of obtaining $P \leq 0.025$ in the first real study.

If we assume that the original studies in the Open Science Collaboration had a planned power of 80\%, then this probability of 0.367 would have been the expected proportion observed with $P \leq 0.025$ in the replication studies. In the event, the observed proportion was 36.1\%. The average P values observed among the 35 out of 97 studies with $P \leq 0.05$ (two-sided) would have been subject to stochastic variation (the 95\% confidence interval for 35/97 = 36.1\% was 26.6\% to 46.2\%), so the estimate of 36.7\% was surprisingly close.

An interesting situation arises with respect to the demand of regulatory bodies for two study results with P values of $≤ 0.05$ two-sided or $≤ 0.025$ one-sided. Thus, if the first study's P value is $0.025$ one-sided, the probability of a second study of the same size getting a significant result is only 0.283. In order to get a 0.8 probability of getting a significant result again when P = $0.025$, we would need a second sample size of just over 4 times the first. If 4 separate studies of the same size were done, there is a probability of 0.283 of getting a significant result after the 1st study, 0.5 after the 2nd study, 0.67 after the 3rd study and 0.8 after the 4th study. This only applies on average before any repeat studies are done. However, if we combined results cumulatively, then the probability of replication will vary stochastically as more data comes in, the variation reducing with more data.

Many trialists are aware that conventional power calculations underestimate the sample sizes required to achieve significant P values as frequently as desired. This has led to other ways of getting better success rates such as a 'mathematical multiplier' (e.g., by setting a statistical power of 95\% to suggest greater sample sizes) if funding permits. Other approaches include Bayesian Assurance, and setting an extremely large sample size arbitrarily and adopting strict rules in the protocol allowing them to look at the data after an interval to allow them to stop if a satisfactory end point has been achieved.  

A power calculation is based on the likelihood of detecting an estimated point value of the true effect size and could be termed the 'likelihood' or 'statistical' power calculation. However, the probability of replicating the thought experiment with the same estimated effect size and same estimated standard deviation, to achieve a probability of replication of 0.8 in the planned real study, could be termed the 'predictive' power calculation.

\subsection{Parallel RCTs}

The above examples involve cross-over RCTs. The same probabilities of replication would apply to parallel-design RCTs, but the sample sizes would of course need to be approximately twice the sample size in each limb compared with that required for a cross-over RCT (a total of about four times the sample size for a cross-over RCT—see the bottom row of Table~2). For a parallel-design RCT, $b$ is the raw mean difference between groups (as opposed to the within-subject difference for a cross-over RCT). The standard deviation $s$ is the standard deviation pooled across groups for parallel RCTs (as opposed to within-subject differences for cross-over RCTs). $n_1$ and $n_2$ reflect the number of independent observations contributing to each estimate per group in parallel RCTs (but per subject in cross-over RCTs). Equation~5, which gives the probability of replication conditional on the P value in the original study, applies automatically to both cross-over and parallel RCT designs.

\section{Comparison with another expression for replication success}
Another expression that estimates the probability of replication success that follows on from Goodman's original paper [9] is:

\[
P\left( z_{\text{repl}} > 1.96\frac{s}{\sqrt{n_2}} \mid b, s, n_1, n_2 \right)
 = \Phi\left(
 \frac{b - 1.96 \frac{s}{\sqrt{n_2}}}{
 \sqrt{\frac{s^2}{n_1} + \frac{s^2}{n_2}}}
 \right)
\tag{Equation 14}
\]

Equation~14 can be modified as follows

\[
P\left( z_{\text{repl}} > 1.96\frac{s}{\sqrt{n_2}} \mid b, s, n_1, n_2 \right)
 =  \Phi\left(
 \frac{b}{
 \sqrt{\frac{s^2}{n_1} + \frac{s^2}{n_2}}}
 - \frac{1.96 \frac{s}{\sqrt{n_2}}}{
 \sqrt{\frac{s^2}{n_1} + \frac{s^2}{n_2}}}
 \right)
\tag{Equation 15}
\]

When $P_1$ is the one-sided P value of the completed study, $P_2$ is the future one-sided P value of the second, replicating study, $P_3$ is the maximum desired one-sided P value of the replicating study (in this case, $P_3 = 0.025$ when $ \Phi^{-1}(P_3) = -1.96$) and the sample sizes are the same in the original and replicating studies (i.e., $n_1 = n_2)$, Equation~15 simplifies to the following Equation 16 (see Appendix 3):

\[
P(P_2 \leq P_3 \mid P_1)
 =\Phi\left( \frac{\Phi^{-1}(P_3) - \Phi^{-1}(P_1)}{\sqrt{2}}
 \right) \tag{Equation 16}
\]
\\
The new expression described in this paper that also aims to estimate the probability of replication (like Equations 14 and 15) is Equation 1:
\[
p(P_2 \leq P_3 \mid b, s, n_1, n_2)
 = \Phi\left(
 \frac{b}{\sqrt{ \left( \frac{s}{\sqrt{n_1}} \right)^2
 + \left( \frac{s}{\sqrt{n_2}} \right)^2 }}
 + \Phi^{-1}(P_3)
 \right)
\tag{Equation 1}
\]
Equation 1 can also be rearranged and simplified to give Equation 5 (see Appendix~2) (which is analogous to Equation 16):

\[
p(P_2 \leq P_3 \mid P_1)
 = \Phi\left(\Phi^{-1}(P_3) -\Phi^{-1}(P_1)/\sqrt{2}\right)
\tag{Equation 5}
\]
\\
Van Zwet and Goodman~[7] have pointed out that the approaches represented by Equations~14, 15 and 16 are overoptimistic about a replication reaching significance (i.e. they are poorly calibrated). Thus for an observed $P = 0.001$ (a Z value of 3.29 in Figure~3), the predictive power estimated from the Cochrane Database is 64\% (see the highest upper-right black marker in Figure~3). This is exactly what Equation~5 (corresponding to the green line in Figure~3) predicts; the same applies to most of the other black markers in Figure~3 from the Cochrane Database, except for the two lowest markers. However, Equation~16 (corresponding to the red line in Figure~3) suggests a probability of 0.83 at a Z value of 3.29, which—as van Zwet and Goodman point out—is a substantial overestimate [7]. The red diamond marker in Figure~3 represents the 36.1\% frequency of replication from the Open Science Collaboration study, which is predicted closely by Equation~5. In contrast to Equation 16, Equation 5 is well calibrated with respect to the data from the Open Science Collaboration study and the Cochrane data base.

\begin{figure}
        \centering
        \includegraphics[width=1.0\linewidth]{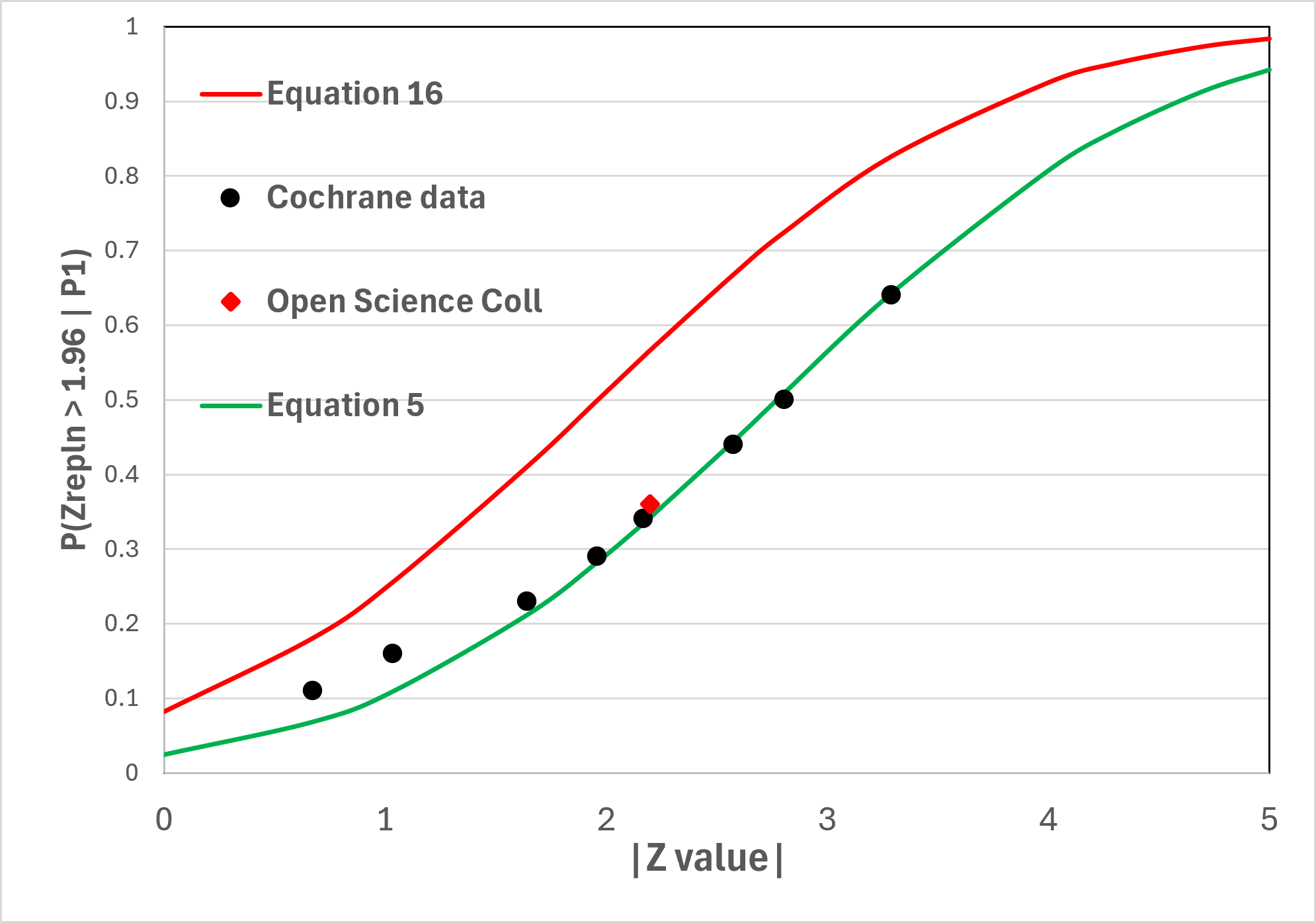}
        \caption{Comparison of various equations for predicting observed frequencies of replication from the Cochrane Database (circular black markers) and the Open Science Collaboration study (red diamond marker)}
        \label{fig:placeholder}
    \end{figure}

\section{Implications of results so far}

Equation~5 is based on three main assumptions. The first is that the studies were designed properly with appropriate statistical modelling and conducted impeccably, with no scope for bias, etc. The second assumption is that  prior probability is a sound model for random sampling. The third assumption is that the result of replicating a study can be modelled by adding the variances of the first and second study in a process of convolution. It is a testament to those who conducted the studies analysed in this paper that the observed frequencies of replication from the Cochrane Database and Open Science Collaboration were as expected from these idealistic assumptions.

The assumption of uniform prior probabilities applies to the random sampling model and not to other prior probabilities—especially those encountered in clinical practice, for example, or Bayesian prior probabilities that might modify the probability of replication. The assumption of uniform prior probabilities also implies that 95\% confidence intervals from random sampling can be interpreted as indicating a probability of 0.95 that the true result will lie within the interval if a “random sampling” study were continued until there was a near-infinite number of samples, conditional on the data obtained so far at each point. It is still open for this probability, conditional on the data alone, to be modified by a Bayesian prior distribution. The probability of replication with $P \leq 0.025$ conditional on a similar sample size gives an indication of how that probability will vary on its journey toward a theoretically infinite sample size as more data are obtained. It is hoped that these concepts may help prevent students from being baffled by P values and confidence intervals.

The foregoing arguments also suggest that one should distinguish between two types of power calculation:

1. Statistical power, which is the likelihood (e.g.\ 80\%) of a study detecting some specified true effect size with a P value of up to 0.05 two-sided or 0.025 one-sided, given an estimated standard deviation and sample size.

2. Predictive power, which is the probability of seeing a study result with a P value of up to 0.05 two-sided or 0.025 one-sided, given an estimated effect size, estimated standard deviation, and specified sample size. Typically, twice the sample size is required to achieve a predictive power (or probability) of 0.8 compared with that required for a statistical power of 80\%.

\section{Investigating whether the apparent true result is due to bias or other sources of distortion}

The probability (e.g.\ 0.975) of a true mean being within a range of interest (e.g.\ a difference of over 0\%) only suggests that the result of the study is promising. This could be interpreted as stating that the probability of it being the true result is up to 0.975 (i.e.\ anywhere between 0 and 0.975). However, this result may not be due to the phenomenon of interest (e.g.\ the treatment being more effective than placebo). It could be due to bias, dishonesty, cherry-picking, data dredging, or other forms of distortion. Each of these possible explanations can be investigated by looking for items of evidence that make the invalid causes unlikely, thereby leaving a genuine treatment effect as the most probable explanation. Mayo (2018) calls this process \emph{severe testing}~[14]. This is analogous to a physician working logically through a list of possible spurious explanations for a diagnostic finding before addressing the differential diagnoses of diseases that might be amenable to treatment.

The probability theory for investigating spurious causes of findings—or rival diagnoses or scientific hypotheses—can be modelled using a derivation of Bayes’ rule with a dependence assumption~[15]. This derivation was formulated to model the differential diagnostic process as performed by physicians, which can be described as a probabilistic version of the disjunctive syllogism. It uses ratios of probabilities between pairs of rival diagnoses (or rival scientific hypotheses) in addition to ratios of likelihoods between them. It can also consider the probability of some other explanation not yet considered (so-called “abductive” reasoning). Failure to apply such \emph{severe testing}, as also suggested by Mayo (2018), may contribute to the low frequency of replication currently observed. If severe testing fails to discount a source of distortion in the results, then adjustments can be made to compensate for it using Bayesian methods.

\section{Testing scientific hypotheses}

We may not only be interested in the effectiveness of a treatment (e.g.\ an angiotensin receptor blocker (ARB) being better than placebo) but also in the theories about the underlying mechanisms. For example, if we had already found evidence supporting the efficacy of an ARB in preventing nephropathy, we might wish to conduct another study comparing the efficacy of an ARB with a non-ARB that lowers blood pressure to the same extent but does not theoretically block the angiotensin receptor or reduce protein leakage. If the ARB were shown to be more effective than the non-ARB in this situation, then this would support the hypothesis that the ARB’s additional effect was due to reducing renal protein leakage, in addition to lowering systemic blood pressure. The result would not confirm the angiotensin-receptor theory because another, as yet unconsidered, explanation might also be compatible with the findings. This is consistent with Karl Popper’s assertion that hypotheses cannot be proven, but that rival hypotheses can be refuted (or at least shown to be less probable).

\section{Conclusion}
Conventional statistical practice evaluates evidence by reference to a null hypothesis, typically through P values derived from the sampling distribution of a single study. In contrast, the present approach focuses on the predictive distribution of replication outcomes conditional on the observed estimate. This shift in perspective leads naturally to standardisation based on the combined uncertainty of both studies rather than emphasising the uncertainty in the replicating study alone. As a result, this model provides a direct, well-calibrated estimate of the probability that a future study will yield a result exceeding a specified threshold, which is more aligned with the practical concerns of researchers and clinicians. This was achieved by directly estimating outcome probabilities based on observed frequencies, discretising continuous values, assuming uniform prior probabilities, and adding the variances of the original and replicating studies. The fact that the resulting equations predicted observed replication frequencies accurately suggests that these assumptions were realistic.

If the P value is based on a null hypothesis of zero and the second study sample size is infinitely large, then, based on the above assumptions, the corresponding probability of same-sign replication with respect to zero is $1{-}P$, where {P} is one-sided. This provides another interpretation of the $P$ value. However, if the replicating study has the same sample size as the original study, the probability of same-sign replication is lower than ${1-}P$ one-sided. Furthermore, to estimate the probability of replication by obtaining a one-sided P value of 0.025 or less again, we need a same-sign result that is also 1.96 SEMs greater than zero. The probability of replication would then be lower still, of the order found in empirical replication studies. Thus, a replication probability is a function of the estimated effect size of the original study and the combined variance obtained by convolution of the original and replicating study distributions.

It is also important to combine by convolution the variance of an effect size estimate with the expected equal variance in the results of the planned study when it is being carried out. This effectively doubles the planned sample size required to achieve a specified level of statistical significance compared with conventional power calculations. 

\section*{Competing interests}
No competing interests are declared.

\section*{Author contributions statement}
HL conceived the work, wrote and reviewed the manuscript.

\section*{Acknowledgments}
I thank discussants on the DataMethods discussion group for their valuable comments.

\section*{References}

\noindent 1. Spiegelhalter D. \textit{The Art of Statistics: Learning from Data}. Penguin Random House, 2019, p.~305. \\

\noindent 2. Ibid., p.~241. \\

\noindent 3. Wasserstein RL, Lazar NA. The ASA's Statement on P values: Context, Process, and Purpose. \textit{The American Statistician}. 2016;70(2):129–133. \\

\noindent 4. Greenland S, Senn SJ, Rothman KJ, Carlin JB, Poole C, Goodman SN, Altman DG. Statistical Tests, P values, Confidence Intervals, and Power: A Guide to Misinterpretations. Online Supplement to the ASA Statement on Statistical Significance and P values. \textit{The American Statistician}. 2016;70(2):129–133. \\

\noindent 5. Benjamin DJ, Berger JO, Johannesson M, Nosek BA, Wagenmakers E-J, Berk R, et al. Redefine statistical significance. \textit{Nature Human Behaviour}. 2018;2(1):6–10. doi:10.1038/s41562-017-0189-z \\

\noindent 6. Open Science Collaboration. Estimating the reproducibility of psychological science. \textit{Science}. 2015;349(6251):aac4716. \\

\noindent 7. van Zwet EW, Goodman SN. How large should the next study be? Predictive power and sample size requirements for replication studies. \textit{Statistics in Medicine}. 2022;41(16):3090–3101. doi:10.1002/sim.9406 \\

\noindent 8. van Zwet E, Gelman A, Greenland S, Imbens G, Schwab S, Goodman SN. A new look at P values for randomized clinical trials. \textit{NEJM Evidence}. 2023;3(1):EVIDoa2300003. \\

\noindent 9. Goodman S. A comment on replication, P‐values and evidence. \textit{Statistics in Medicine}. 1992;11(7):875–879. PMID:1604067 \\

\noindent 10. Klein RA, Vianello M, Hasselman F, et al. Many Labs 2: Investigating Variation in Replicability Across Samples and Settings. \textit{Advances in Methods and Practices in Psychological Science}. 2018;1(4):443–490. doi:10.1177/2515245918810225 \\

\noindent 11. Senn S. A comment on replication, P values and evidence by S.N. Goodman. \textit{Statistics in Medicine}. 2002;21:2437–2444. doi:10.1002/sim.1072 \\

\noindent 12. Killeen PR. An alternative to null-hypothesis significance tests. \textit{Psychological Science}. 2005;16:345–353. \\

\noindent 13. Llewelyn H. Replacing P values with frequentist posterior probabilities of replication—when possible parameter values must have uniform marginal prior probabilities. \textit{PLOS ONE}. 2019;14(2):e0212302. \\

\noindent 14. Mayo D. \textit{Statistical Inference as Severe Testing: How to Get Beyond the Statistics Wars}. Cambridge University Press; 2018. \\

\noindent 15. Llewelyn H, Ang AH, Lewis K, Abdullah A. \textit{The Oxford Handbook of Clinical Diagnosis}. 3rd ed. Oxford University Press; 2014, pp.~638–642. \\

\newpage
\section*{Appendix 1}

\noindent \textbf{Proof that}

\[
p(P_2 \leq P_3 \mid b, s, n_1, n_2)
= \Phi\!\left(
\frac{b}{\sqrt{\left(\frac{s}{\sqrt{n_1}}\right)^2 + \left(\frac{s}{\sqrt{n_2}}\right)^2}}
+ \Phi^{-1}(P_3)
\right)
\]

When $P_1$ is the one-sided P value from the first completed study (e.g.\ 0.025), $P_2$ is the one-sided P value of the planned replicating study. It is unknown because it refers to a future result; we can only estimate the probability of it being less than or equal to some threshold (e.g.\ $\leq 0.025$). $P_3$ is the desired threshold of the maximum one-sided P value in the replication study. $b$ is the observed raw effect size in the first study, $s$ is the standard deviation, $n_1$ the sample size of the original study, and $n_2$ the planned sample size of the replication study.

Assume a flat prior on $\theta$, and that
\[
b \mid \theta, v_1 \sim N(\theta, v_1),
\qquad
b_{\mathrm{repl}} \mid \theta, v_2 \sim N(\theta, v_2),
\]
where $b$ and $b_{\mathrm{repl}}$ are conditionally independent given $\theta$.

1. Because $b \mid \theta, v_1 \sim N(\theta, v_1)$ and $\theta$ has a flat prior, it follows that  
\[
\theta \mid b, v_1 \sim N(b, v_1).
\]

2. Because $b_{\mathrm{repl}} \mid \theta, v_2 \sim N(\theta, v_2)$ and the two statistics are conditionally independent given $\theta$,  
\[
b_{\mathrm{repl}} \mid b, v_1, v_2 \sim N(b,\, v_1 + v_2).
\]

3. Standardise the predictive distribution by dividing $b_{\mathrm{repl}}$ by $\sqrt{v_1 + v_2}$.

Replication success on a two-sided 0.05 scale corresponds to
\[
P(z_{\mathrm{repl}} > 1.96 \mid b, v_1, v_2)
= P\!\left( \frac{b_{\mathrm{repl}}}{\sqrt{v_1 + v_2}} > 1.96 \right).
\]

4. Conditionally on $b, v_1, v_2$,
\[
\frac{b_{\mathrm{repl}}}{\sqrt{v_1 + v_2}}
\sim N\!\left(
\frac{b}{\sqrt{v_1 + v_2}},\, 1
\right).
\]

Hence,
\[
P(z_{\mathrm{repl}} > 1.96 \mid b, v_1, v_2)
= \Phi\!\left(
\frac{b}{\sqrt{v_1 + v_2}} - 1.96
\right).
\]

Finally, substituting  
\[
v_1 = \left(\frac{s}{\sqrt{n_1}}\right)^2,
\qquad
v_2 = \left(\frac{s}{\sqrt{n_2}}\right)^2,
\]
gives
\[
P(Z_{\text{repl}} > 1.96 \mid b, s, n_1, n_2)
= \Phi\!\left(
\frac{b}{
\sqrt{
\left( \frac{s}{\sqrt{n_1}} \right)^2 +
\left( \frac{s}{\sqrt{n_2}} \right)^2
}}
- 1.96
\right).
\]

Replacing $-1.96$ with $\Phi^{-1}(P_3)$ yields the general expression:

\[
p(P_2 \leq P_3 \mid b, s, n_1, n_2)
= \Phi\!\left(
\frac{b}{
\sqrt{
\left( \frac{s}{\sqrt{n_1}} \right)^2 +
\left( \frac{s}{\sqrt{n_2}} \right)^2
}}
+ \Phi^{-1}(P_3)
\right).
\]

\section*{Appendix 2}

\noindent \textbf{Proof that}

\[
P(P_2 \leq P_3 \mid P_1)
 = \Phi\left(\Phi^{-1}(P_3) -\Phi^{-1}(P_1)/\sqrt{2}\right)
\tag{Equation 5}
\]

Let $z_1$ be the original $z$-statistic, so that
\[
z_1 = \frac{b}{s/\sqrt{n_1}} = \frac{b\sqrt{n_1}}{s}.
\]

Starting from
\[
P(Z_{\text{repl}} > 1.96 \mid b, s, n_1, n_2)
= \Phi\!\left(
\frac{b}{
\sqrt{
\left(\frac{s}{\sqrt{n_1}}\right)^2
+
\left(\frac{s}{\sqrt{n_2}}\right)^2}}
- 1.96
\right),
\]
rearranging yields
\[
P(Z_{\text{repl}} > 1.96 \mid b, s, n_1, n_2)
= \Phi\!\left(
\frac{b}{
\frac{s}{\sqrt{n_1}}
\sqrt{1 + \frac{n_1}{n_2}}}
- 1.96
\right).
\]

When $n_1 = n_2$,
\[
P(Z_{\text{repl}} > 1.96 \mid b, s, n_1, n_2)
= \Phi\!\left(
\frac{b}{(s/\sqrt{n_1})\sqrt{2}}
- 1.96
\right).
\]

Using the relationship
\[
\frac{b}{s/\sqrt{n_1}} = -\Phi^{-1}(P_1),
\]
we obtain
\[
P(Z_{\text{repl}} > 1.96 \mid P_1)
= \Phi\!\left(
-\frac{\Phi^{-1}(P_1)}{\sqrt{2}}
- 1.96
\right).
\]

Replacing $1.96$ with $-\Phi^{-1}(0.025)$ gives
\[
P(Z_{\text{repl}} > 1.96 \mid P_1)
= \Phi\!\left(
-\frac{\Phi^{-1}(P_1)}{\sqrt{2}}
+ \Phi^{-1}(0.025)
\right).
\]

Thus for a general threshold $P_3$,
\[
P(P_2 \leq P_3 \mid P_1)
= \Phi\!\left(
-\frac{\Phi^{-1}(P_1)}{\sqrt{2}}
+ \Phi^{-1}(P_3)
\right).
\]

Rearranging:

\[
P(P_2 \leq P_3 \mid P_1)
 = \Phi\left(\Phi^{-1}(P_3) -\Phi^{-1}(P_1)/\sqrt{2}\right)
\tag{Equation 5}
\]

\section*{Appendix 3}

\[
P\left( z_{\text{repl}} > 1.96\frac{s}{\sqrt{n_2}} \mid b, s, n_1, n_2 \right)
 =  \Phi\left(
 \frac{b}{
 \sqrt{\frac{s^2}{n_1} + \frac{s^2}{n_2}}}
 - \frac{1.96 \frac{s}{\sqrt{n_2}}}{
 \sqrt{\frac{s^2}{n_1} + \frac{s^2}{n_2}}}
 \right)
\tag{Equation 15}
\]
\paragraph{Equation 15 simplifies to:}
\
\[
P\left( z_{\text{repl}} > 1.96\frac{s}{\sqrt{n_2}} \mid b, s, n_1, n_2 \right)
 =  \Phi \left( \frac{b}{\frac{s}{\sqrt{n_1}}\sqrt{2}} - \frac{1.96 \cdot \left(\frac{s}{\sqrt{n_2}}\right)}{\frac{s}{\sqrt{n_1}}\sqrt{2}} \right)
\]

\paragraph{If $n_1=n_2$, the common standard-error term can be factored out and the expression simplifies to:}
\[
P(P_2 \leq P_3 \mid P_1)
= \Phi \left( \frac{\frac{b}{\frac{s}{\sqrt{n_1}}
{}} - 1.96}{\sqrt{2}} \right)
\]
Using the relationship
\[
\frac{b}{s/\sqrt{n_1}} = -\Phi^{-1}(P_1),
\]
and replacing $-1.96$ with $\Phi^{-1}(P_3)$ 
we obtain
\[
P(P_2 \leq P_3 \mid P_1)
 =\Phi\left( \frac{-\Phi^{-1}(P_1) + \Phi^{-1}(P_3)}{\sqrt{2}}
 \right) 
\]
 Rearranging we get 

 \[
P(P_2 \leq P_3 \mid P_1)
 =\Phi\left( \frac{\Phi^{-1}(P_3) - \Phi^{-1}(P_1)}{\sqrt{2}}
 \right) \tag{Equation 16}
\]

\end{document}